# Inflation Determinants in Argentina (2004-2022)[1]

Pablo de la Vega[2]

Guido Zack[3]

Jimena Calvo[4]

Emiliano Libman[5]

This version: November 22, 2023

**Abstract**

This paper analyzes the empirical relationship between the inflation rate and its proximate determinants in Argentina, using quarterly data over the period 2004-2022 and an error-correction vector model approach. Unlike previous literature, this paper uses a theoretical framework to motivate the inclusion of variables that are expected to contribute to explain inflation, thus reducing the risk of omitting relevant variables and formalizing key mechanisms. Inference is performed through Granger causality analysis, impulse response functions and forecast errors variance decomposition. The results suggest that an anti-inflationary plan for Argentina should take into consideration both the greater relevance of the inertial component, the exchange rate and the interest rate in the short-run dynamics of the price level, and the long-run relationship between prices, interest rate and activity level.

[1] We thank Fernando Toledo, Fernando Morra, and the participants of the VI Jornadas Argentinas de Econometría and the LVII Reunión Anual de la Asociación Argentina de Economía Política for their comments. Any errors are our sole responsibility.
[2] Fundar and Instituto de Investigaciones Económicas, Facultad de Ciencias Económicas, Universidad Nacional de La Plata, Argentina. Corresponding author: delavegapc@gmail.com.
[3] Fundar, IIEP (UBA–CONICET), and CIMaD (EEyN-UNSAM).
[4] Fundar.
[5] CONICET and Fundar.

# 1. Introduction

Argentina's inflation levels have shown an increasing trend over the last two decades, recently reaching levels of around 100% per year (Figure 1). There is a relative consensus that these levels are detrimental to economic growth, income distribution and poverty. Thus, inflation is possibly the most relevant problem in Argentina today. However, there are great differences regarding what are the determinants of this inflationary process (Zack et al., 2017). This is crucial, since the anti-inflationary policies to be applied depend on the understanding of the causes and mechanisms that generate the generalized and sustained increase in the price level.

**Figure 1. Monthly year-over-year inflation (%), 2004-2022**

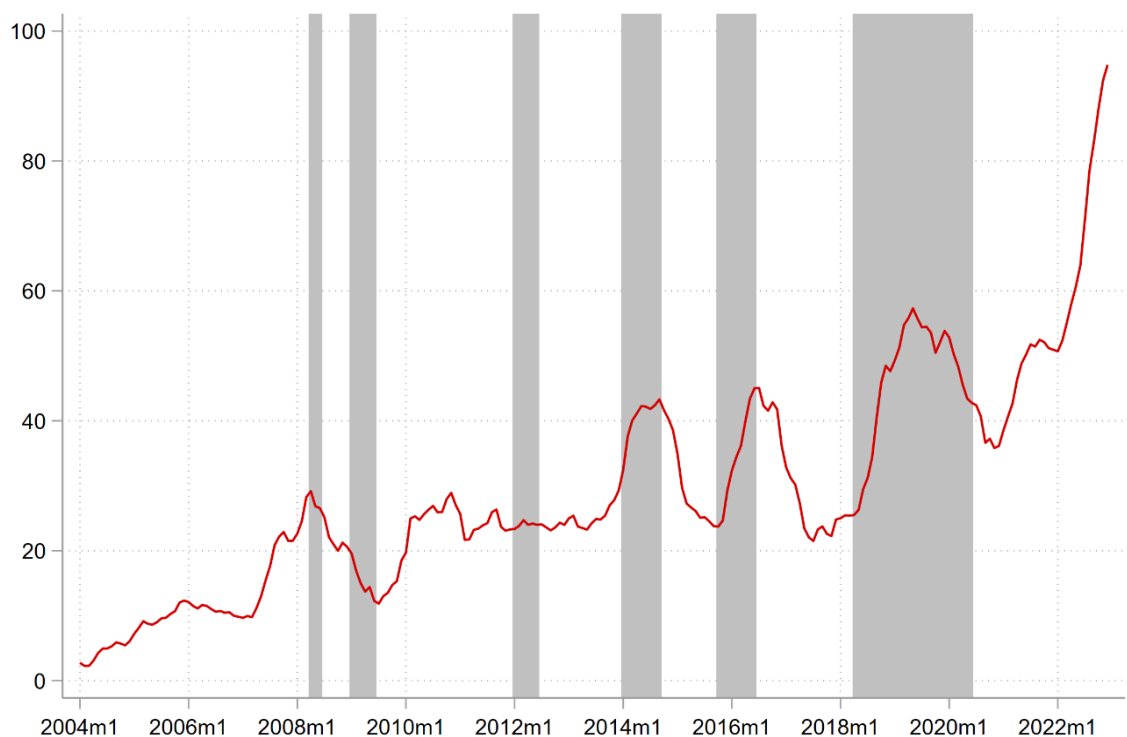

Own elaboration based on INDEC and Provincial Statistics Institutes. The figure shows the year-on-year variation of the consumer price index. The gray area indicates periods in which there has been an economic contraction for two consecutive quarters.

This paper contributes to the literature on the determinants of inflation in Argentina by using quarterly data for the period 2004-2022 and a vector error correction (VEC) model approach, which allows us to analyze both long-run relationships and short-run dynamics between variables that are determined simultaneously. Unlike previous literature, this paper starts from a theoretical accounting scheme that decomposes the price level into its proximate determinants and then motivates the inclusion of different variables that are expected to contribute to explain inflation. This reduces the risk of omitting relevant variables and, at the same time, formalizes key mechanisms. Inference is performed through Granger causality analysis, impulse response functions (IRFs) and forecast error variance decomposition (FEVD) under different Cholesky decompositions.

The results suggest that, in the long run, the price level has a positive relationship with the interest rate. In addition, the price of regulated goods and the international price of imports are positively associated with the price level, but these relationships are not statistically significant. Similarly, the exchange rate is negatively associated with the price level, but this relationship is not statistically significant. These results are consistent with an economy that has suffered stagflation for a significant part of the sample. Interest rate changes in an environment of instability seem to fuel inflationary processes (e.g., because they fuel inflation expectations or raise firms' cost of capital).

We test for weak exogeneity, analyzing which of the variables respond to deviations from the estimated long-run relationships. The nominal exchange rate and international commodity prices are found to be weakly exogenous, suggesting that they are the candidate variables to be the common trends driving the system. The adjustment of the remaining system variables to the long-run relationships demonstrates the empirical value of using a systemic approach to obtain consistent estimates.

We then concentrate on analyzing the response of the price level to exogenous shocks in the rest of the system variables by means of IRFs and FEVDs. The results are, in general, as expected. Shocks to the price level have a positive and permanent impact on the price level itself. The price of regulated goods has a positive impact in the first quarter, but becomes non-significant from the second quarter onwards, with a decreasing trend; this seems to suggest that tariff increases are inflationary in the short term, due to their direct impact, but the effect is diluted in the long run (and could even contribute to reduce inflation if it allows balancing public accounts). Nominal depreciations have a positive and lasting impact over time on the price level. The activity level does not have a statistically significant effect in any of the orders. Money supply shocks only have a positive effect in the first quarter in orders 1 and 2, where this variable is at the beginning of the recursive sequence. The most controversial result is the positive effect of interest rate shocks on the price level. This result is common in the literature, which calls it the "price puzzle" for being counterintuitive in theoretical terms (Sims, 1986; Castelnuovo and Surico, 2009; Rusnak et al., 2013; Estrella, 2015). Finally, the price of imports does not have a relevant effect on the price level.

The variance decomposition shows the importance of the shocks to each variable in explaining the variations of the variables in the model and how this importance varies over time. For example, between 50 and 58% of the variation in prices in the initial period is due to shocks at the same price level. However, this importance decreases over time and it is between 16 and 28% in the sixth quarter, so that the contribution of other variables becomes more relevant. In orders 1 and 2, the variable whose contribution grows the most is the interest rate, which goes from explaining 24% of the variance in the first period to 48% in the sixth. Something similar, but to a lesser extent, occurs with the exchange rate, which goes from explaining 15% in the first period to 25% in the last period. On the other hand, in orderings 3 and 4, the exchange rate is the variable whose contribution increases the most over time, going from explaining 33% to 56%. The increase in the contribution of the interest rate is considerable but smaller under these orders. Money supply is only relevant under orders 1 and 4, although its contribution does not exceed 8% in any period. Similarly, the contribution of the price of international goods does not exceed 8% in any period, but its contribution is present in all four orders. The price of regulated goods has a contribution in the first two quarters of between 4 and 8%. Finally, the level of activity does not have a relevant contribution in any of the simulations.

The rest of the paper is organized as follows. The next section reviews the previous literature on the determinants of inflation in Argentina and establishes a theoretical framework that motivates the econometric specification used in the rest of the paper. Secction 3 describes the data and the empirical methodology, the results of which are discussed in Section 4. Conclusions are included in Section 5.

## 2. Literature review

The discussion on the origins and the problem of combating inflation has been the subject of an extensive literature, in many cases motivated by the Latin American experience. This literature has several edges, among which the discussion on the origins of inflationary processes, the different types of regimes (which differ according to their duration, intensity and sustainability over time) and the types of strategies used to control them in those cases in which prices increase at a detrimental rate.

Broadly speaking, there are three hypotheses on the origin of inflation: a) monetary and fiscal (Sargent, 1982); b) distributional conflict (Lavoie, 2022, chapter 8) and; c) structural (Heymann, 1986). Although the bulk of the profession and the literature is inclined to explain the phenomenon on the basis of the first explanation, there has recently been a resurgence of interest in approaches associated with the second and third interpretations (Werning and Lorenzoni, 2023).

The distinction is not always clear-cut. For example, it is possible to observe an interrelation between distributive conflict and macroeconomic policy. An expansive fiscal policy that results in unsustainable situations may have as a background the inability of society to set a budget that can be financed with taxes and without explosive indebtedness (Heymann and Navajas, 1989; Gerchunoff and Rapetti, 2016).

In addition, in low inflation contexts, the connection between the activity level, money and prices is not so clear. Based on the criticism of the "Phillips Curve" (Phillips, 1958) Friedman (1968) and Phelps (1967) suggested the hypothesis that in the long run unemployment is at its "natural rate", revealing the "monetary" character of inflation. The Friedman and Phelps hypothesis assumes that deviations from the natural rate generate changes in the inflation rate, something that does not seem to be particularly observed when inflation is above the natural rate (Akerlof et al., 2000)[6].

Conflict inflation approaches often introduce "structural" issues (Olivera, 1991). For example, it is usually assumed that the supply of industrial goods adjusts to demand (given the existence of excess capacity), but the supply of food is inelastic in the short run (Cardoso, 1981). Thus, changes in demand require changes in relative prices that are transformed into inflation if the authorities validate subsequent wage increases or if they introduce subsidies to keep food prices low (Agenor and Montiel, 2008).

---

[6] Other papers have suggested that the presence of "supply shocks" explains the tendency of the "natural rate" to move over time, constituting a sort of moving target that makes its identification difficult in practice (Gordon, 2013). Others have suggested that the presence of "path dependence" implies that the "observed" rate of unemployment modifies the "natural rate"; in this case, contractionary policies have permanent effects on unemployment (Blanchard and Summers, 1986).

On the contrary, in hyperinflation contexts it seems to be easier to find some consensus on the primacy of the monetary issue. The model par excellence of hyperinflation is that of Cagan (1956), but authors such as Kalecki (1962) suggested a very similar formulation.

In summary, it is essential to analyze inflation as a non-linear phenomenon, which can be classified according to the specific macroeconomic conditions. A distinction to be considered is that between "chronic inflation" and "high inflation and hyperinflationary" regimes. In a context of "chronic inflation" (Pazos, 1972), price and wage dynamics exhibit a marked inertia, so that past inflation becomes a guide to predict price dynamics.

In addition, prices tend to adjust at different times of the year. In a context of "high inflation and hyperinflation", past inflation rates do not provide a reliable guide for making economic decisions and adjustments are made more frequently and synchronously, so that exchange rate movements are most likely to be taken as a reference (Heymann and Leijonhufvud, 1995).

It is precisely the lack of a price index in local currency as a mechanism to guide decisions that makes the difference between "high inflation" and "hyperinflation". The short duration of contracts during hyperinflation causes the mere reduction in the rate at which prices grow to raise real wages, which are set with longer lags than prices. For this reason, when hyperinflation stops, an expansion of the level of activity is usually observed (Taylor, 1991). This mechanism is absent in stabilization experiences in non-hyperinflationary contexts of high inflation.

The literature on stabilization experiences has taken note of the differences that exist not only among the different inflationary regimes, but also in the variety of approaches to combat inflation. Consequently, it classifies plans according to their most salient features. For example, a distinction is usually made between "orthodox" programs, which emphasize the correction of macroeconomic imbalances and macroeconomic policy inconsistencies (fiscal deficit, exchange rate overvaluation, etc.), and "heterodox" programs, which focus on the need to intercede in the price and wage formation process (to combat inflationary inertia). It is generally recognized that no stabilization attempt has had lasting success without combining a dose of both programs (Dornbusch and Simonsen, 1987).

Another classification explores the distinction between stabilization programs based on regulating the quantity of money and stabilization programs based on the use of a fixed or semi-fixed exchange rate. In Latin America and in undeveloped countries, programs based on exchange rate stabilization tend to predominate. Reinhart and Végh (1994) analyze seventeen stabilization programs (for Argentina, Brazil, Dominican Republic, Israel, Mexico, Peru and Uruguay), of which five were based on the quantity of money and the remaining twelve on exchange rate anchors. This work includes a series of stylized facts that analyze the average evolution of macroeconomic aggregates before, during and after the implementation of the stabilization plan. Perhaps the most important stylized fact is the one that shows that stabilization with an exchange rate anchor first produces an expansion of output and then a recession, while monetary stabilization reverses the order.

The empirical study of the determinants of inflation has received substantial attention in the literature, particularly in countries that have experienced sustained inflationary phenomena over time such as Argentina (Chhibber, 1991; Akinboade et al., 2001, Helmy, 2008; Ndikumana et al., 2021). A striking point in that literature is that most papers estimate an econometric specification that is not derived from a theoretical model, but is a compilation of variables that, according to previous literature, would be expected to have an effect on the price level (Dhakal et al., 1994; Kim, 1998; Khan and Schimmelpfennig, 2006; Tran, 2018; Lakshmanasamy, 2022). However, there are certain papers that do start from a formalized theoretical scheme to motivate empirical analysis (Chhibber, 1991; Akinboade et al., 2001; Lissovolik, 2003; Nguyen and Niguyen, 2010; Nguyen et al., 2012; Akinbobola, 2012; Elgammal and Mohamed, 2016).

In the case of Argentina, there are several papers that analyze the same period as the present article and with similar methodologies (Zack et al., 2017; Graña Colella, 2020; García-Cicco et al., 2022). These works share the use of time series econometrics, in particular, of cointegration vector models, which is particularly necessary in the case of the analysis of variables that are determined simultaneously and, in addition, evidence a long-term relationship between them.

Zack et al. (2017) estimate two VEC models to describe the dynamics of the consumer price level from October 2004 to February 2016. The first model includes the money supply and the level of activity while the second one adds, in addition, the level of wages and the exchange rate. After obtaining the long-run relationships, the authors analyze IRFs and FEVDs assuming two different Cholesky decompositions. They find that inertia and the exchange rate are the most important factors in explaining inflation. Graña Colella (2020) also uses a VEC model, but using quarterly data for the period 2003-2019. Based on the literature review, the author includes international prices, unit labor cost, nominal exchange rate and money supply as explanatory variables. His results suggest that unit labor cost and the exchange rate are key to explain the price level in the long run, but in the short run monetary issuance and inertia play a role.

Montes Rojas and Toledo (2021) use a VAR model with directional quintiles to estimate the inflationary impact of shocks on the international price of agricultural commodities and on the exchange rate. They find that the first shock generates a 10% pass-through, while the second one generates a 25% pass-through. Inspired by the post-Keynesian-structuralist literature (Vera, 2014, Abeles and Panigo, 2015), the authors point out the relevance of the distributional conflict resulting from the fall in the real wage as a transmission mechanism to the price level.

García-Cicco et al. (2022) study the stylized facts of inflationary processes for a panel of Latin American countries including Argentina. The variables they incorporate in the system are the core price index, the exchange rate, the interest rate, the level of activity, the money supply, a wage index, the output gap and international food and energy prices. Although they also perform a cointegration analysis to study the long-run relationships between variables, they decompose the evolution of year-on-year inflation based on local projections (Jordà, 2005). In the long run they find that, in Argentina, the price level is related to the wage index and the exchange rate. Finally, they find that inertia and exchange rate movements are the main explanatory factors of the inflationary process in Argentina. Monetary and activity variables play a minor role and only in specific periods.

Our paper builds on Zack et al. (2017), extending the analysis in at least three directions. First, we establish a theoretical framework that decomposes the price level into its proximate determinants and motivates the inclusion of several variables in the econometric specification, which allows us to reduce the risk of omitting relevant variables, formalize key mechanisms, and establish the assumptions on which the analysis is developed.[7] Second, the set of simulations is extended by considering different orderings with respect to the contemporaneous relationship between variables in order to provide greater robustness to the results. Finally, we include not only domestic variables but also external variables such as the price of international commodities and extend the period of analysis to July 2022.

---

[7] The chosen methodology has the ability to account for the simultaneity in the determination of the variables included in the analysis. However, this does not completely rule out the possibility of omitting relevant variables and, therefore, the presence of biases.

## 2.1 A theoretical framework on the decomposition of the price level[8]

Without loss of generality, the general price level ($P$) can be expressed as a weighted average of the price of tradable ($P^T$), non-tradable ($P^N$) and regulated ($P^R$) goods. Taking logarithms (in lower case letters) we have the following:

$$p = \theta_0 p^N + \theta_1 p^T + \theta_2 p^R \qquad (1)$$

with $\theta_0 + \theta_1 + \theta_2 = 1$

In a small open economy (international price taker), purchasing power parity[9] holds and $p^T$ can be expressed in domestic currency as a function of international prices ($p^f$) and the nominal exchange rate ($er$, domestic currency per U.S. dollar):

$$p^T = er + p^f \qquad (2)$$

The price of non-tradable goods can be explained in at least two different ways (Lissovolik, 2003). On the one hand, assuming that the non-tradable goods market has the same trend as the aggregate goods market, $p^N$ can be expressed as a function of domestic money supply and demand (Lissovolik, 2003, Nguyen et al., 2012; Akinbobola, 2012; Elgammal and Mohamed, 2015):

$$p^N = \beta(m^s - m^d) \qquad (3)$$

where $m^s$ y $m^d$ are the supply and demand for real balances, respectively; and $\beta$ is a parameter. It is also usual to assume that $m^d$ is a function of real income ($y$), the expected inflation rate ($\pi^e$) and the interest rate ($i$) as follows[10]:

$$m^d = m - p = f(\underset{+}{y}, \underset{-}{\pi^e}, \underset{+/-}{i}) \qquad (4)$$

The arguments of the demand for real balances denote the different motives for which non-remunerated and remunerated money is demanded. Thus, for example, a higher level of activity induces a higher transactional demand for money, so that a higher demand for goods correlates with a higher demand for money. Conversely, higher inflationary expectations decrease the demand for money and increase the demand for goods, since money holdings are expected to buy fewer goods in the next period. Meanwhile, a higher interest rate on domestic currency assets decreases both the demand for goods and non-interest-bearing money but increases the demand for interest-bearing money. Therefore, the expected effect of the interest rate is, in principle, ambiguous. However, a fall in the expected real interest rate decreases the demand for money (remunerated and non-remunerated) and increases the demand for goods.

---

[8] The following section closely follows Chhibber (1991), Akinboade et al. (2001), Lissovolik (2003), Nguyen et al. (2010), Nguyen et al. (2012), Akinbobola (2012), and Elgammal and Mohamed (2015).
[9] Purchasing power parity is the generalization to all goods of the law of one price according to which, in the absence of trade restrictions and frictions, the price of the same good will be the same in any country in the world, in equilibrium.
[10] We implicitly assume that the relevant substitution is between goods and domestic money. In an economy with a high degree of currency substitution, the functional form of money demand could, in addition, consider the role of the exchange rate in portfolio decisions. This becomes even more complex in a context of exchange rate restrictions and the emergence of parallel exchange rates.

The definition of $\pi^e$ will depend on the assumption about the formation of expectations. Generally speaking, it could be assumed either that expectations are formed by looking forward (forward-looking) or backward (backward-looking).

On the contrary, other theoretical developments choose to explain $p^N$ in terms of a markup model over costs as follows:

$$p^N = (1 + \mu)(\omega + \sigma) \tag{5}$$

where $\mu$ is a markup coefficient, $\omega$ is the unit labor cost, $\sigma$ is the cost of intermediate goods (both domestic and imported, so that $\sigma$ is a function of $er$ and $p^f$).

Then, we have two price equations that can be estimated empirically. From (1)-(4) we have the following price equation:

$$p = f(\underset{+}{m^s},\ \underset{+/-}{y}\ ,\underset{+}{\pi^e},\underset{-}{i},\underset{+}{er},\underset{+}{p^f},\underset{+}{p^r}) \tag{6}$$

On the other hand, from (1), (2) and (5), it follows that:

$$p = f(\underset{+}{\mu},\underset{+}{\omega},\underset{+}{\sigma}\ ,\underset{+}{er},\underset{+}{p^f},\underset{+}{p^r}) \tag{7}$$

The signs below each variable indicate the expected effect on $p$ according to the equations presented above.[11] Note that the last three determinants are the same in both models and identify some of the usual causes of inflation: imported inflation ($p^f$), cost inflation due to exchange rate changes ($er$), and inflation due to changes in regulated prices ($p^r$)[12]. On the other hand, the first four determinants in (6) are associated with what is known as demand inflation, while the first three determinants in (7) refer to the effect of labor costs, intermediate input costs and changes in profit margins on inflation.

---

[11] Note that the coefficient of $y$ can be negative due to money demand considerations (see equations (3) and (4)), but positive for reasons associated with the output gap.

[12] Although it is determined by the government, it is expected to depend on variables included in the model, such as the exchange rate. In turn, the policy decisions associated with its determination have strong fiscal and monetary implications and, therefore, an indirect effect on prices.

The variables to be included broadly coincide with those considered by previous literature (García-Cicco et al., 2022). However, Zack et al. (2017) and Graña Colella (2020) do not include the interest rate, nor the price of regulated goods and services. Zack et al. (2017), moreover, do not include international prices.

It is important to bear in mind that this theoretical framework is not exhaustive in that it does not trace the ultimate causes of the inflationary process, but only the proximate ones. For example, the scheme does not include the fiscal deficit among the explanatory factors since it does not have a direct impact on prices, but it does affect several of the variables included in the analysis, such as the money supply, the interest rate, the exchange rate, among others. In this sense, for some authors, the fiscal deficit is the ultimate cause of inflation (García-Cicco, 2021).

On the other hand, the post-Keynesian-structuralist tradition (Vera, 2014, Abeles and Panigo, 2015) emphasizes the distinction between the causes of inflation and the transmission mechanisms (Montes Rojas and Toledo, 2021). This literature points to the relevance of distributional conflict as a transmission mechanism of supply shocks such as those in commodity prices, which lower real wages and trigger a potentially destabilizing race between prices and wages.

## 3. Data and Methodology

Empirical analysis using quarterly data[13] for the period from January 2004 to July 2022 of the variables identified in [Section 2.1](). More specifically, we will use the variables that best capture the evolution of the money supply, the exchange rate, the interest rate, inflation expectations, wages and the different "prices" (tariffs, commodities, etc.) that enter as arguments of the theoretical model.

Due to data availability constraints, it is necessary to assume that inflation expectations are formed in a backward-looking manner, so that $\pi^e$ refers to past inflation.[14] In any case, this does not seem to be a very strong assumption for this period in Argentina. It is also necessary to assume that the markup coefficient remains constant due to the absence of data in this respect. All series are used seasonally adjusted and measured in logarithms, except for the interest rate, which is expressed as a percentage of one. Table 1 describes each variable and its source of information, while Table 2 shows descriptive statistics.

---

[13] The results are virtually the same when using monthly frequencies and they are available upon reasonable request.
[14] Although there is a series of inflation expectations that could be constructed from the BCRA's Relevamiento de Expectativas de Mercado, it does not have sufficient time coverage.

**Table 1. Variables and sources of information**

| Name (associated variable of Section 2.1) | Description | Source |
|---|---|---|
| CPI | Consumer Price Index (2012m10=100) | INDEC, Provincial Statistical Institutes |
| M2 | M2 Monetary Aggregate | BCRA |
| Activity Level | Monthly Estimator of Economic Activity (EMAE) (2012m10=100) | INDEC, Provincial Statistical Institutes |
| Interest Rate | Badlar, deposits of more than $ 1M | BCRA |
| NEER | Nominal Effective Exchange Rate Index (2012m10=100). | BCRA |
| Imports Prices | Price Index of 45 Imported Commodities including agricultural commodities, energy, food and beverages, and metals (2012m10=100) | IMF |
| Regulated Prices | Regulated Price Index (2012m10=100) | IIEP (UBA), IPCBA, INDEC |
| Wage Index | Wage Index (2012m10=100) | INDEC |

Notes: Own elaboration.

**Table 2. Descriptive statistics**

|  | N | Mean | SD | Min | p25 | p50 | p75 | Max |
|---|---|---|---|---|---|---|---|---|
| CPI | 75 | 5.00 | 1.40 | 3.16 | 3.78 | 4.73 | 6.05 | 7.93 |
| M2 | 75 | 13.19 | 1.38 | 11.08 | 11.99 | 13.10 | 14.29 | 16.01 |
| Activity Level | 75 | 4.52 | 0.12 | 4.19 | 4.47 | 4.57 | 4.60 | 4.64 |
| Interest Rate | 75 | 0.20 | 0.13 | 0.02 | 0.10 | 0.18 | 0.27 | 0.54 |
| NEER | 75 | 5.15 | 1.08 | 3.92 | 4.26 | 4.69 | 5.79 | 7.45 |
| Imports Prices | 75 | 4.35 | 0.30 | 3.62 | 4.10 | 4.31 | 4.62 | 5.05 |
| Regulated Prices | 75 | 5.35 | 1.23 | 4.13 | 4.31 | 4.76 | 6.40 | 7.95 |
| Wage Index | 75 | 4.90 | 1.35 | 2.96 | 3.71 | 4.74 | 5.99 | 7.62 |

Notes: Own elaboration.

One way of analyzing the relationship between simultaneously determined economic variables in a non-structural manner is to estimate econometric models such as the vector autoregressive model (VAR) and the vector error correction model (VEC) (Johansen, 1988; Johansen and Juselius, 1990; Juselius, 2006; Lütkepohl, 2007).

In a VAR model, each variable is considered as a function of the past values of all the variables in the system. However, when there is a cointegrating relationship between the variables in the system, the VAR is misspecified and the correct method is the VEC model, which is a VAR model with cointegrating restrictions (Engle and Granger, 1987). The formal specification of the latter is as follows:

$$\Delta y_t = \alpha\beta' y_{t-1} + \sum_{i=1}^{l-1} Y_i \Delta y_{t-i} + \mu_t \tag{8}$$

where $y_t$ is a time series vector, $y_t \sim I(d)$, $l$ is the number of lags of the underlying VAR, $\mu_t$ is the error term, and $\beta' y_{t_{t-1}}$ is the error correction term reflecting the long-run equilibrium relationship between the variables.[15] The matrix $\beta$ characterizes the cointegrating relationships between the variables, while $\alpha$ indicates the speed of adjustment to an imbalance with respect to this long-run relationship. In this way, the VEC model allows us to analyze both the long-run equilibrium relationship between the variables, as well as the short-run disequilibrium. In order to perform some kind of structural inference, some assumptions are made about the ordering of the variables (following, for example, the Cholesky decomposition).

## 4. Results

The two models formulated in the theoretical framework of Section 2.1 were estimated, however, the model given by equation (7) did not allow obtaining good diagnostic tests and failed in the stability tests. Therefore, the results shown below only summarize the findings for the model given by equation (6).

### 4.1 Unit root tests

The Dickey-Fuller test is used to analyze the level of integration of the variables. The results shown in Table 2 suggest that all the variables are integrated of order 1. In other words, the null hypothesis of unit root is not rejected in levels, but it is rejected in first differences.

---

[15] Since the price series is used in logarithms, for small changes we have that $\Delta p_t = \Delta \ln(P_t) = \ln(P_t) - \ln(P_{t-1}) \approx \pi_t$, where $\pi_t$ is the inflation rate between t and t-1.

**Table 2. Unit Root Tests**

|  | Augmented Dickey-Fuller | | Deterministic components |
|---|---|---|---|
|  | Level | Difference |  |
| CPI | 0.996 | 0.001 | constant and trend |
| M2 | 0.996 | 0.015 | constant and trend |
| Activity Level | 0.364 | 0.006 | constant and trend |
| Interest Rate | 0.241 | 0.000 | constant |
| NEER | 0.982 | 0.013 | constant and trend |
| Imports Prices | 0.239 | 0.044 | constant |
| Regulated Prices | 0.881 | 0.005 | constant and trend |

Notes: The table shows the p-values associated with the null hypothesis that the series has a unit root.

## 4.2 Cointegration test

Since the series are integrated of order 1, we proceed to perform the cointegration test, the results of which are presented in Table 3.[16] The null hypothesis of no cointegration is rejected, while the null hypotheses of the existence of three and four cointegrating relationships at 1% and 5% significance, respectively, are not rejected.

**Table 3. Johansen Cointegration Test (Trace Test)**

| Range | statistic | 5% | 1% |
|---|---|---|---|
| 0 | 193.0178 | 146.76 | 158.49 |
| 1 | 136.1557 | 114.90 | 124.75 |
| 2 | 97.0854 | 87.31 | 96.58 |
| 3 | 67.5869 | 62.99 | 70.05 |
| 4 | 41.7409 | 42.44 | 48.45 |
| 5 | 21.5913 | 25.32 | 30.45 |
| 6 | 5.4194 | 12.25 | 16.26 |
| 7 |  |  |  |

Notes: A restricted trend is included in the cointegration space.

## 4.3 VEC model estimation

As already mentioned, the VEC model allows us to analyze both the long-run equilibrium relationship between the variables, as well as the short-run disequilibrium. The cointegrating relationships are presented in Tabla 4.[17] Then, the Johansen normalization procedure on the parameters defined, in addition to a price equation, a real stock demand equation.[18]

---

[16] Previously it was determined, based on information criteria, that the optimal number of lags is two.
[17] Although the cointegration test suggests the presence of three or four cointegrating relationships, these models fail the stability tests, so it was decided to estimate only two cointegrating relationships.
[18] The identification of the parameters in the cointegration equation requires the imposition of certain restrictions so that some of them remain fixed.

**Table 4. Long-run relationships**

|  | Eq1 | Eq2 |
|---|---|---|
| CPI | 1.000 |  |
| M2 |  | 1.000 |
| Activity Level | 1.881 | -0.616 |
|  | (0.805)** | (3.627) |
| Interest Rate | -3.902 | -14.411 |
|  | (0.629)*** | (2.835)*** |
| NEER | 0.154 | -1.612 |
|  | (0.297) | (1.337) |
| Imports Prices | -0.024 | 0.482 |
|  | (0.152) | (0.685) |
| Regulated Prices | -0.004 | 2.026 |
|  | (0.207) | (0.931)** |

Notes: The standard deviations of the estimated coefficients are shown in parentheses. * $p<0.1$; ** $p<0.05$; *** $p<0.01$.

In the long run, the price level has a negative and statistically significant relationship with the level of activity, while it has a positive relationship with the interest rate.[19] In addition, the price of regulated goods and the international price of imports are positively associated with the price level, but these relationships are not statistically significant. Similarly, the exchange rate is negatively associated with the price level, but this relationship is not statistically significant. According to the second cointegration relationship, the demand for money balances is positively associated with the interest rate and negatively associated with the price of regulated. In addition, the exchange rate and the level of activity are positively related to the price level, but these associations are not statistically significant. Finally, the international price of imports is positively associated with the M2 aggregate, but this relationship is also not statistically significant.

Although there is no single way to interpret these results, the estimates are consistent with the existence of contractionary effects associated with accelerating inflation. On the other hand, upward movements in interest rates, since they are associated with increases in the inflation rate, are consistent with the existence of traditional monetary policy transmission mechanisms to prices. For example, a rise in interest rates can increase inflation expectations or raise the cost of working capital of companies, raising the inflation rate.

---

[19] Since the variables of the system, with the exception of the interest rate, are expressed in logarithms, the coefficients can be interpreted as elasticities. For example, in the long run, a 1% increase in the level of activity is associated with a 1.881% fall in the price level.

The coefficient of the error correction term of the first cointegrating relationship in the price equation is 0.072, which implies that the system corrects the previous period's disequilibrium at a rate of 7.15% per quarter. In other words, it will take 3.5 years to reach equilibrium in the face of an unexpected shock.

W weak exogeneity tests are performed, analyzing which of the variables respond to deviations from the estimated long-run relationships (Johansen, 1992; Juselius, 2006). As shown in Table 5, the nominal exchange rate and the price of international commodities are weakly exogenous. This means that these variables do not respond to deviations from long-run relationships, suggesting that they are the candidate variables to be the common trends that move the system. The adjustemnt of the rest of the variables in the system to the long-run relationships demonstrates the empirical value of using a systems approach to obtain consistent estimates.[20]

**Table 5. Weak exogeneity test**

|  | chi2 | p-value |
|---|---|---|
| CPI | 4.645 | 0.098 |
| M2 | 8.309 | 0.016 |
| Activity Level | 6.500 | 0.039 |
| Interest Rate | 5.075 | 0.079 |
| NEER | 1.450 | 0.484 |
| Imports Prices | 2.613 | 0.271 |
| Regulated Prices | 12.443 | 0.002 |

Notes: The table shows the chi2 statistic and p-values associated with the weak exogeneity test for each variable in the model.

Given the complexity of the interactions between the variables of the system, the analysis of the short-run relationship between the variables is usually performed, not through point estimates of the coefficients, but through Granger causality tests, IRFs and FEVDs. However, before performing such exercises, it is advisable to perform diagnostic tests to elucidate whether the model performs well.

---

[20] Strictly speaking, this does not prevent weakly exogenous variables from not responding to long-run relationships, but their adjustment is less immediate than endogenous ones.

## 4.4 Diagnostic tests

Inference from the model requires that the cointegration equations be stationary. As shown in Figure 2, the remaining eigenvalues[21] are within the unit circle, although there is one that is 0.98. Although there is no theory to determine how far away from 1 they must be for the model to be stable, the cointegration equations are stationary, as shown in Figure 3. Likewise, the model errors do not show serial autocorrelation (see Table 6).

**Figure 2. Stability of the VEC model**

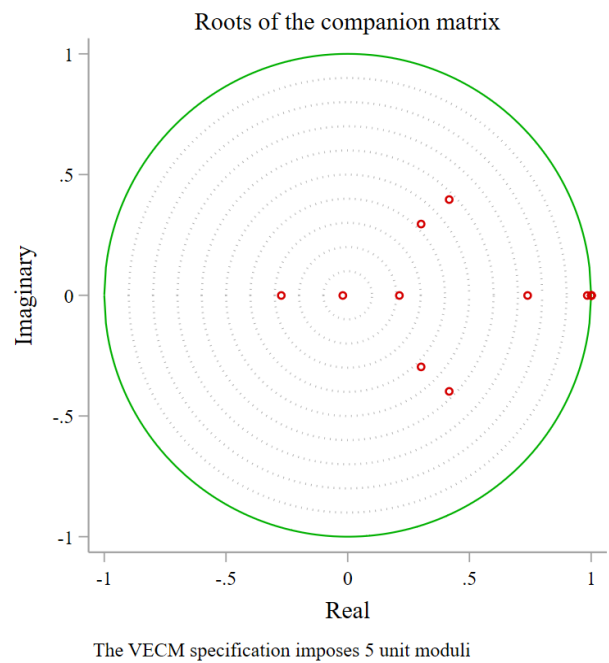

Own elaboration.

**Figure 3. Cointegration relationships**
Panel A: Cointegration relationship 1

---

[21] If the VEC model has K variables and r cointegrating vectors, there will be K-r unit modules in the matrix. In our case, K=7 and r=2, so K-r=5.

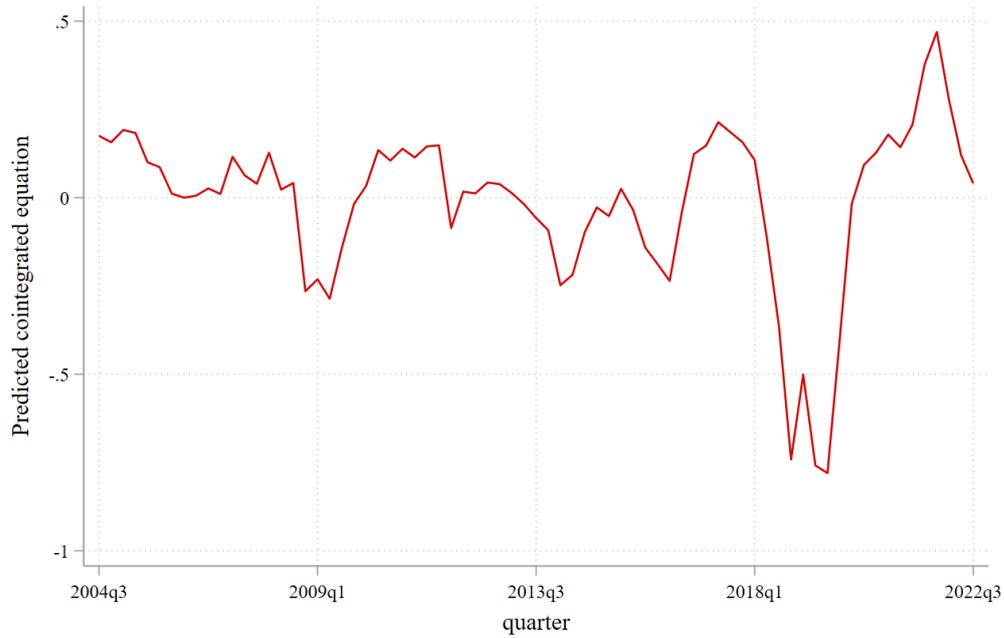

Panel B: Cointegration relationship 2

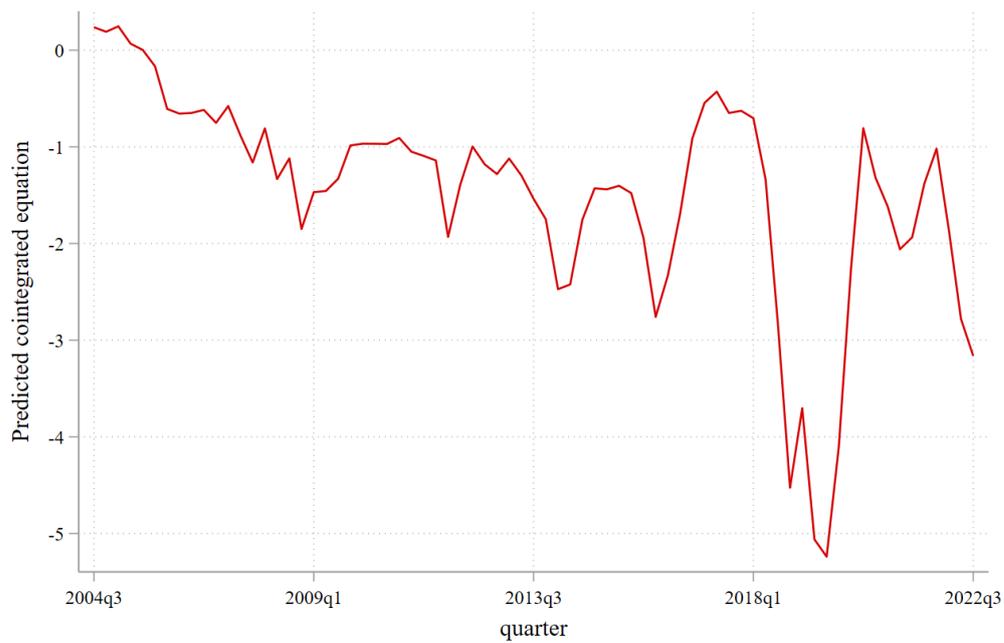

Own elaboration.

**Table 6. Autocorrelation test**

| Lag | chi2 | df | Prob>Chi2 |
|---|---|---|---|
| 1 | 37.74 | 49 | 0.879 |
| 2 | 49.326 | 49 | 0.46 |

Notes: The table shows the results for the Lagrange multiplier test, whose null hypothesis is the absence of serial correlation.

## 4.5 Inference

The first inference exercise consists of analyzing temporal precedence relationships between pairs of variables, which is usually called causality in the Granger sense (Granger, 1969). The intuition of this test is simple. Variable A Granger causes variable B if past changes in A help predict B. If such a temporal precedence relationship exists, it can be unidirectional or bidirectional.[22] In the specific context of a VEC model, this test consists of evaluating the null hypothesis that all lags of variable A in the equation of variable B are zero, so that we refer to short-run Granger causality.[23] This is particularly relevant because in a cointegration context the existence of some kind of Granger causality in the short run is necessary for the system to correct towards long-run equilibrium (Engle and Granger, 1987).

The results of the Granger causality tests are presented in Table 7. The price level is anticipated by changes in the interest rate and the exchange rate. Another expected result is that changes in domestic variables (with the exception of the exchange rate) do not contribute to predict the change in international commodity prices, which supports the assumption that Argentina is an international price taker.

The Activity level does not anticipate any change in the remaining variables of the system, which is logical for international prices, possible for prices (which suggests that changes in activity have little direct effect on inflationary dynamics), exchange rate, regulated prices, interest rates and M2 (variables whose evolution could obey a certain rule of behavior of the authorities). On the other hand, commodity prices and regulated prices anticipate the evolution of the level of activity, which is to be expected, and could be capturing distributive effects.

---

[22] It should be clarified that causality in the Granger sense is strictly associated with the temporal sequence in which the values of the variables are realized, a concept that should be distinguished from that of exogeneity, understood in the more general sense of the term, whereby one variable is the cause of another. In addition, the existence of Granger causality from A to B does not preclude the possibility of the existence of another variable C that also causes B.

[23] It is valid to note that this causality is only in a direct sense in that it does not contemplate the potential relationship of A and B, which is mediated by another variable C.

**Table 7. Short-run Granger causality**

|  | CPI | M2 | Activity Level | Interest Rate | NEER | Imports Prices | Regulated Prices |
|---|---|---|---|---|---|---|---|
| CPI | 0.256 | 0.107 | 0.367 | 0.032 | 0.002 | 0.367 | 0.389 |
| M2 | 0.579 | 0.001 | 0.933 | 0.000 | 0.029 | 0.260 | 0.319 |
| Activity Level | 0.026 | 0.192 | 0.731 | 0.161 | 0.232 | 0.031 | 0.028 |
| Interest Rate | 0.177 | 0.401 | 0.765 | 0.004 | 0.183 | 0.581 | 0.183 |
| NEER | 0.551 | 0.009 | 0.283 | 0.004 | 0.012 | 0.550 | 0.626 |
| Imports Prices | 0.338 | 0.533 | 0.218 | 0.374 | 0.012 | 0.005 | 0.968 |
| Regulated Prices | 0.864 | 0.035 | 0.587 | 0.518 | 0.000 | 0.000 | 0.872 |

Notes: The table shows the p-values on the null hypothesis that the variable in the column does not Granger cause the variable in the row. In green those null hypotheses that are rejected at 5% significance.

In the remainder of this section, we concentrate on analyzing the response of the price level to exogenous shocks in the rest of the system variables by means of IRFs (Figure 4) and FEVDs (Figure 5). Unlike a stationary VAR, the IRFs of a VEC do not necessarily converge to zero over time since the variables are integrated of order 1, which makes it possible to differentiate between permanent and transitory shocks (Lütkepohl, 2007). On the other hand, this analysis requires orthogonalizing the errors of the equations, which is usually done based on the Cholesky decomposition, which implies assuming the recursive contemporaneous relationship between the variables. For example, given three variables A, B and C, the ordering given by A→B→C assumes that A contemporaneously affects B and C, but B and C do not contemporaneously affect A; likewise, B contemporaneously affects C, but not vice versa. In other words, the variables that appear first contemporaneously affect those that follow, but not vice versa. In practice, the ordering usually depends on expected theoretical relationships, while the use of different orderings allows the robustness of the results to be evaluated.

Intuitively, the IRFs allow us to analyze the dynamic effects of shocks to the system variables. In particular, the IRFs presented in Figure 4 show the time response of the consumer price index to an orthogonal shock of one standard deviation in each of the system variables during six quarters under the four different Cholesky orders listed below:

- **Order 1:** Imports Prices → M2 → Interest Rate→ Activity Level → NEER → Regulated Prices → CPI
- **Order 2:** Imports Prices→ M2 → Interest Rate → NEER → Activity Level → Regulated Prices→ CPI
- **Order 3:** Imports Prices → NEER → Regulated Prices→ CPI → Activity Level → M2 → Interest Rate

- **Order 4:** Imports Prices → NEER → Activity Level → Regulated Prices→ M2 → Tasa de interés → CPI

The variables are ordered from most exogenous to least exogenous. Since it is not possible to provide a definitive ranking, the orders reflect some plausible attributes of different economic policy regimes. For example, following the literature on open economies, policymakers may prioritize the use of instruments to regulate exchange rate movements, sacrificing their ability to regulate the interest rate and vice versa. Note, for example, that in orders 1 and 2, the money supply plays a more active role, while in orders 3 and 4 it is more accommodative and the dominant role is played by the nominal effective exchange rate.

Note also that Imports Prices is included first in all four orderings since it is the only exogenous variable in the model (according to Granger's analysis), i.e., it does not depend on domestic conditions. Regulated prices are considered, alternatively, at the beginning and at the end, since it is feasible to consider them an instrument of economic policy (whose objective is to influence the real wage and the inflation rate) but also a by-product of the dynamics of the exchange rate and activity.

First, the IRFs in the four simulations are very similar, which supports the robustness of the results. Second, the results are mostly in line with expectations. Price level shocks have a positive and permanent impact on the price series itself. The price of regulated goods has a positive impact in the first quarter, but becomes non-significant from the second quarter onwards, with a decreasing trend. Nominal depreciations have a positive and lasting impact over time on the price level. The level of activity does not have a statistically significant effect under any of the orders. Money supply shocks have a positive impact only in the first quarter under orders 1 and 2, where this variable is at the beginning of the recursive sequence. The most controversial result is the positive effect of interest rate shocks on the price level. This result is usual in the literature, which denotes it as the "price puzzle" since it is counterintuitive in theoretical terms (Sims, 1986; Castelnuovo and Surico, 2009; Rusnak et al., 2013; Estrella, 2015). Finally, the price of imports does not have a relevant effect on the price level.

**Figure 4. Response of the price level to an orthogonal shock of 1 standard deviation in each variable of the system**
Interest rate

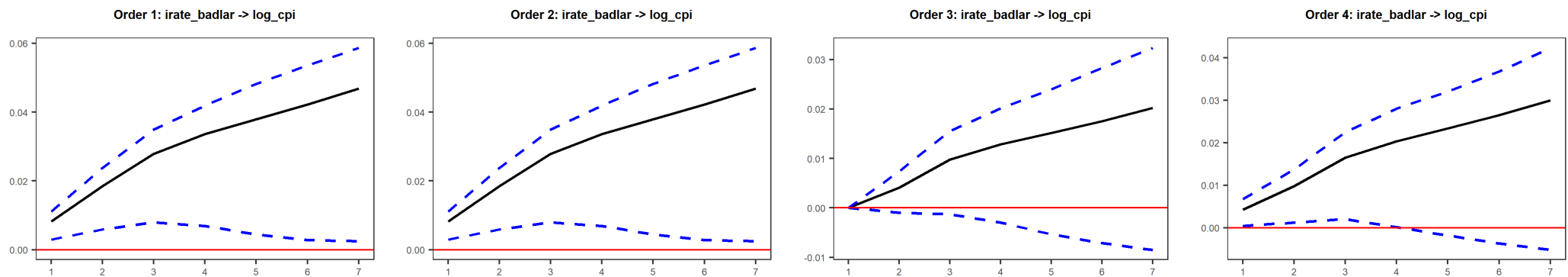

**CPI**

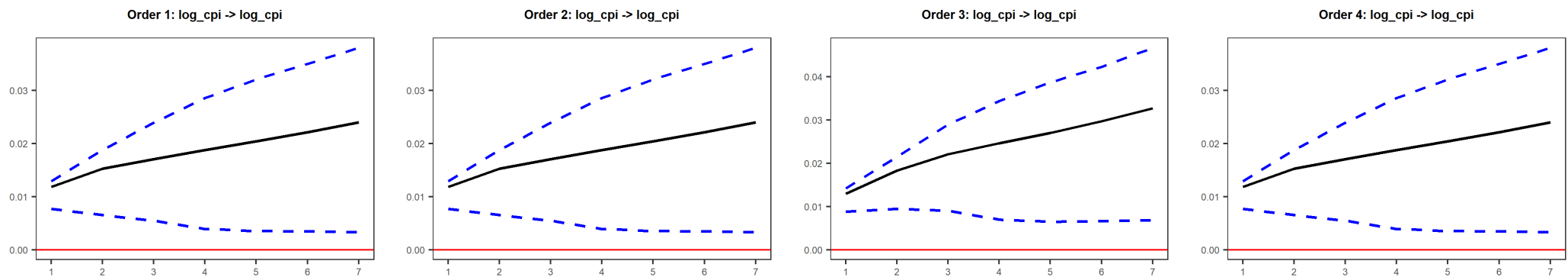

## Activity Level

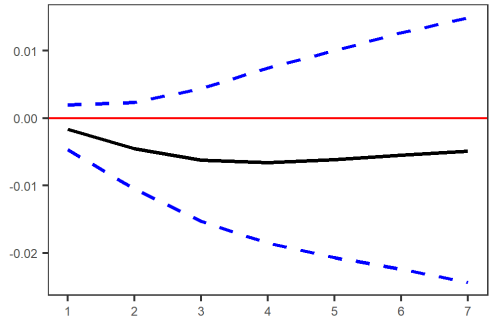
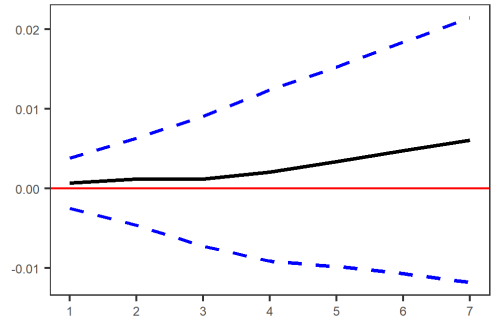
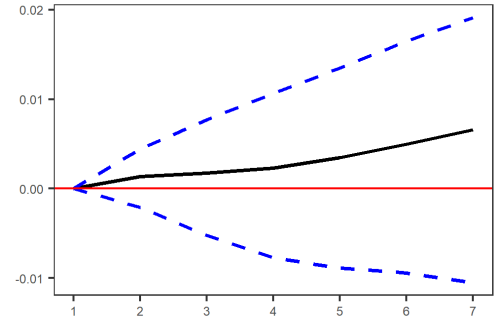
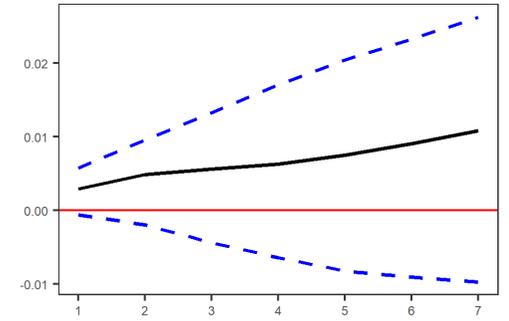

## NEER

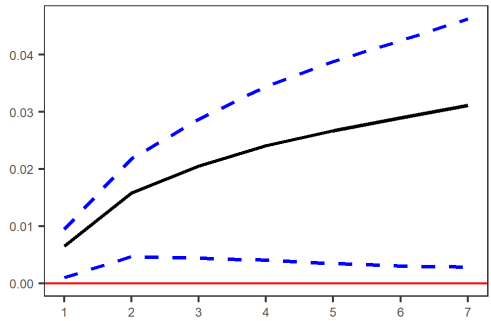
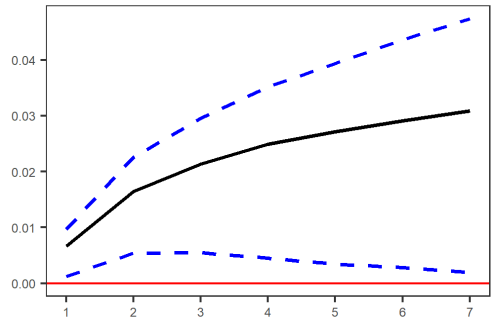
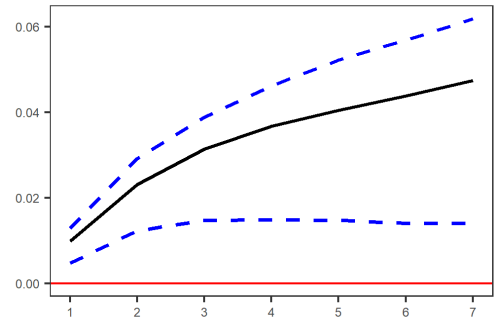
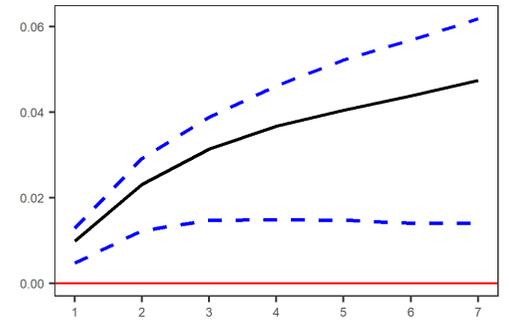

**Imports Prices**

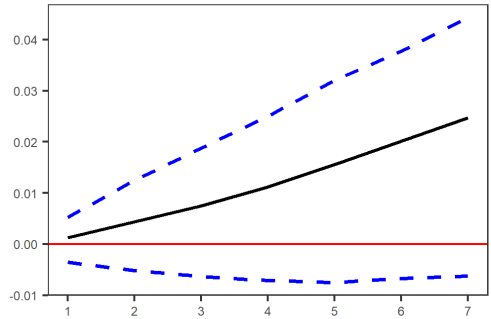 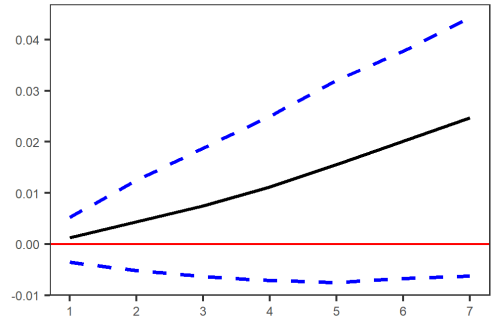 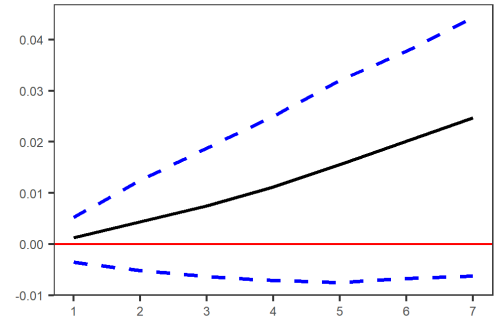 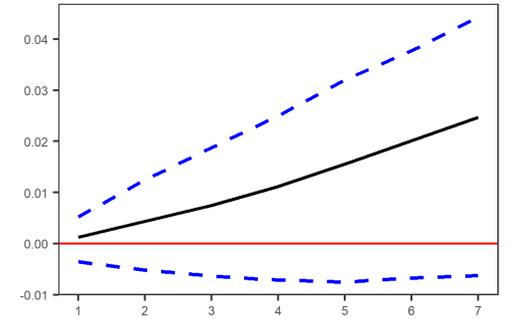

**M2**

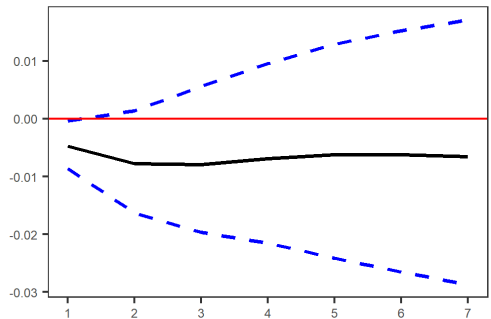 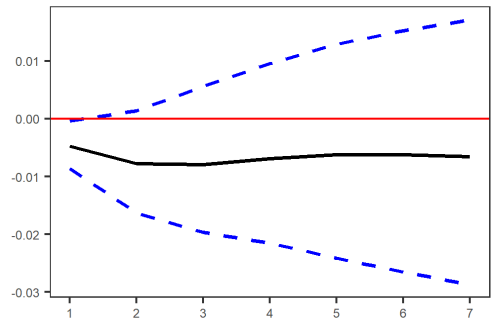 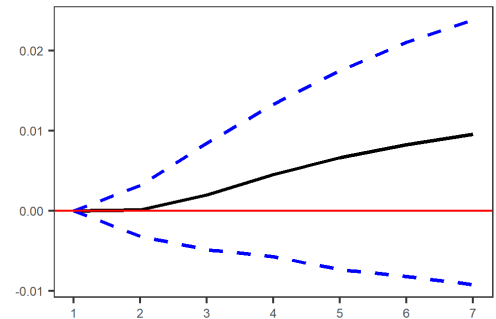 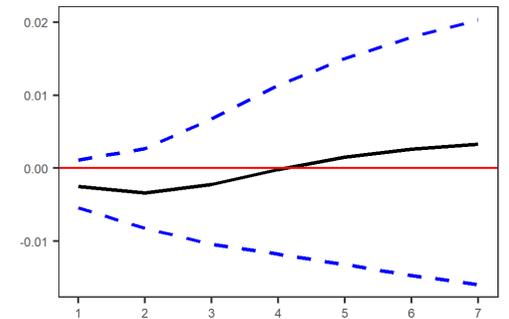

**Regulated Prices**

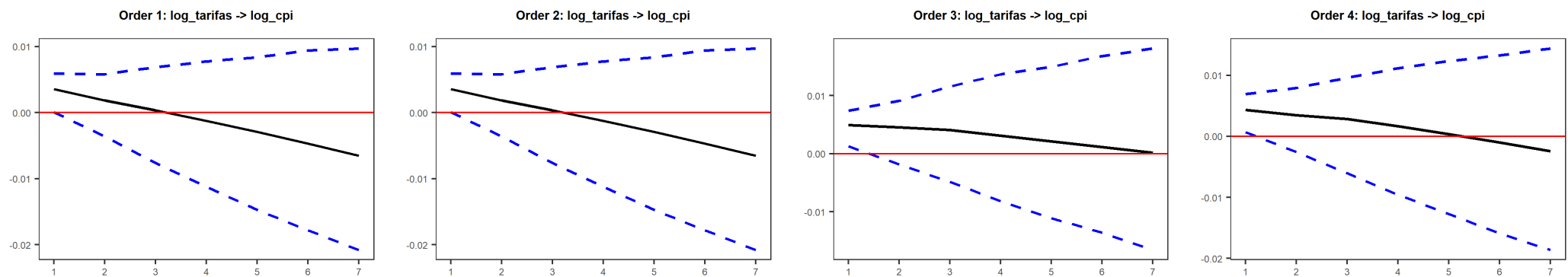

Notes: own elaboration. The figure shows the time response of the consumer price index to an orthogonal shock of one standard deviation in each of the variables of the system during six quarters under different Cholesky orders. 95% confidence intervals obtained by 1000 Bootstrap runs are included. References are as follows: log_cpi = CPI, irate_badlar = Interest Rate; log_emae = Activity Level; log_itcnm = NEER; log_m2 = M2; log_tarifas = Regulated Prices; log_pm = Import Prices. **Order 1:** Imports Prices → M2 → Interest Rate → Activity Level → NEER → Regulated Prices → CPI. **Order 2:** Imports Prices → M2 → Interest Rate → NEER → Activity Level → Regulated Prices → CPI. **Order 3:** Imports Prices → NEER → Regulated Prices → CPI → Activity Level → M2 → Interest Rate. **Order 4:** Imports Prices → NEER → Activity Level → Regulated Prices → M2 → Tasa de interés → CPI.

Finally, [Figure 5](#) shows the analogous results for the variance decomposition. This indicates what proportion of the forecast errors variance of a variable can be attributed to the shocks to the different variables in the system, including itself. Intuitively, it shows how important the shocks are in explaining the variances of the model variables and how this importance varies over time. For example, between 50 and 58% of the variation in prices in the initial period is due to shocks at the same price level. However, this importance falls over time and by the sixth quarter is between 16 and 28%, so that the contribution of other variables becomes more relevant.

In orderings 1 and 2, the variable whose contribution increases the most is the interest rate, which goes from explaining 24% of the variance in the first period to 48% in the sixth. Something similar, but to a lesser extent, occurs with the exchange rate, which goes from explaining 15% in the first period to 25% in the last. On the other hand, in orderings 3 and 4, the exchange rate is the variable whose contribution increases the most over time, going from explaining 33% to 56%. The increase in the contribution of the interest rate is considerable but smaller in these orderings. Money supply is only relevant under orders 1 and 4, although its contribution does not exceed 8% in any period. Similarly, the contribution of the price of international goods does not exceed 8% in any period, but its contribution is present in all four orders. The price of regulated goods has a contribution in the first two quarters of between 4 and 8%. Finally, the level of activity does not have a relevant contribution in any of the simulations.

**Figure 5. Variance decomposition for the CPI according to the different orderings**

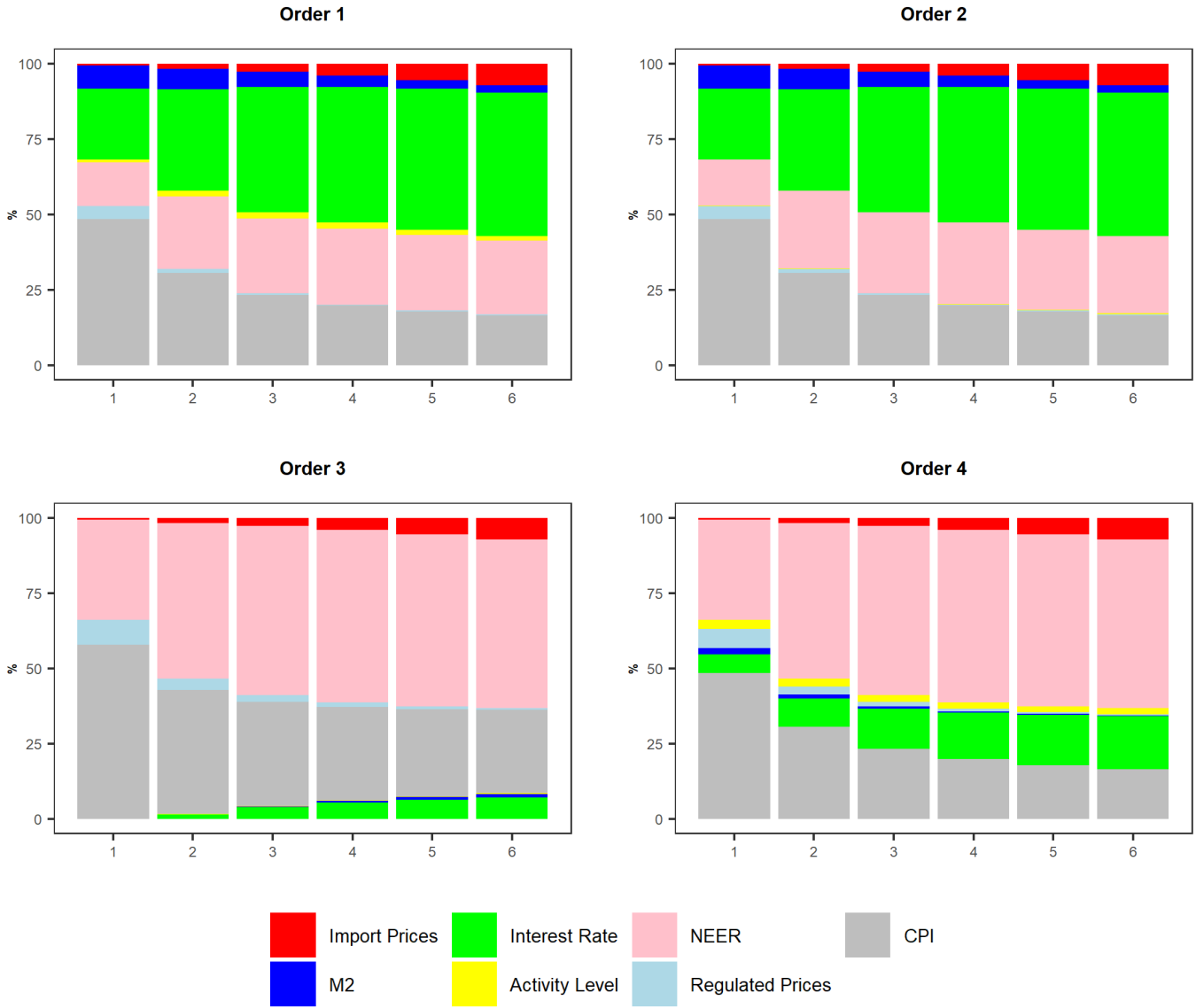

Own elaboration. The figure shows the proportion of the forecast errors variance of a variable that can be attributed to the shocks of the different variables of the system, including itself. **Order 1:** Imports Prices → M2 → Interest Rate→ Activity Level → NEER → Regulated Prices → CPI. **Order 2:** Imports Prices→ M2 → Interest Rate → NEER → Activity Level → Regulated Prices→ CPI. **Order 3:** Imports Prices → NEER → Regulated Prices→ CPI → Activity Level → M2 → Interest Rate. **Order 4:** Imports Prices → NEER → Activity Level → Regulated Prices→ M2 → Tasa de interés → CPI.

# 5. Conclusion

Inflation is one of the most pressing problems facing Argentina, reaching prohibitive levels for growth, improvement in income distribution and poverty reduction. Unfortunately, there are great differences regarding the determining factors of this inflationary process and, therefore, of the anti-inflationary policies to be applied.

In this sense, this paper contributed to the previous literature on the determinants of inflation in Argentina in the last two decades. We used quarterly data during the period 2004-2022 and a VEC model, which allows us to analyze both long-run relationships and short-run dynamics between variables that are determined simultaneously. Unlike previous literature, this paper started from a theoretical scheme that decomposes the price level into its proximate determinants and motivated the inclusion of different variables that are expected to contribute to explain inflation.

Our results suggest that inflation has recessionary effects in the long run, that the interest rate (and more generally monetary policy) seems to have little ability to directly influence inflation in the right direction (i.e. a rate hike does not seem to contribute to reduce inflation, and even seems to accelerate it), except perhaps for its effects on the exchange rate, inflation exhibits considerable inertia (as suggested by the variance decomposition), rate adjustments are inflationary only in the short run, and changes in the level of activity have little capacity to explain the evolution of the inflation rate.

These results indicate that, when designing an anti-inflationary plan for Argentina, both the greater relevance of the inertial component and the inflationary effects of the interest rate, the exchange rate and the prices of regulated products and services (which include, among other things, electricity and gas tariffs) on the short-term dynamics of the price level should be taken into consideration.